\documentclass[12pt,a4paper]{iopart}

\newcommand{\eq}[1]{\begin{equation}#1\end{equation}}
\newcommand{\dd}{\mathrm{d}}
\newcommand{\ee}{\mathrm{e}}

\newcommand{\bea}{\begin{eqnarray}}
\newcommand{\eea}{\end{eqnarray}}
\newcommand{\beas}{\begin{eqnarray*}}
\newcommand{\eeas}{\end{eqnarray*}}
\newcommand{\twovec}[2]{\left(\begin{array}{c} #1 \\ #2 \end{array}\right)}
\newcommand{\twomat}[4]{\left(\begin{array}{cc} #1 & #2 \\ #3 & #4\end{array}\right)}

\newcommand{\Cov}{C}
\newcommand{\unity}{\ensuremath{{\rm 1 \hspace{-0.27778em} l}{}}}

\usepackage{iopams}
\usepackage{graphics}
\usepackage{graphicx}
\usepackage{psfrag}
\usepackage{color}
\usepackage{colordvi}
\usepackage{multibib}

\begin{document}

\title{Entanglement negativity in the harmonic chain out of equilibrium}

\author{Viktor Eisler$^1$ and Zolt\'an Zimbor\'as$^{2,3}$}
\address{$^1$Institute for Theoretical Physics,
E\"otv\"os Lor\'and University, P\'azm\'any s\'et\'any 1/a, H-1117 Budapest, Hungary \\
$^2$Department of Computer Science, University College London, Gower Street,\\ 
{WC1E 6BT} London, United Kingdom\\ 
$^3$Department of Theoretical Physics,
University of the Basque Country UPV/EHU, P.O. Box 644, E-48080 Bilbao, Spain}

\begin{abstract}
We study the entanglement in a chain of harmonic oscillators driven out of equilibrium
by preparing the two sides of the system at different temperatures, and subsequently
joining them together. The steady state is constructed explicitly and the logarithmic negativity
is calculated between two adjacent segments of the chain. We find that, for low temperatures,
the steady-state entanglement is a sum of contributions pertaining to left- and right-moving
excitations emitted from the two reservoirs.
In turn, the steady-state entanglement is a simple average of the Gibbs-state values and thus its
scaling can be obtained from conformal field theory.
A similar averaging behaviour is observed during the entire time evolution.
As a particular case, we also discuss a local quench where both sides of the chain are initialized
in their respective ground states.
\end{abstract}

%\maketitle

\section{Introduction}

In the last decade it has been recognized that the entanglement content of many-body
quantum states carries essential information about the underlying system \cite{Amico08,CCD09}.
In the simplest case, when the system is in a pure state, 
the entanglement between two complementary parts is measured by the entanglement entropy.
One of the most remarkable results for the case of one-dimensional systems is that 
the entanglement entropy of ground states displays a universal behaviour. At critical points
it grows logarithmically in the subsystem size \cite{Vidal03}, with a prefactor governed by the
central charge of the underlying conformal field theory (CFT) \cite{CC09};
while for noncritical chains it saturates to a finite value, i.e., an area law holds \cite{ECP09}.

However, the use of the entropy as an entanglement measure is restricted to pure states  
and a bipartite setting. The entanglement between non-complementary subsystems,
embedded in a larger system, thus needs a different characterization,
since the reduced state is in general mixed.
Among the numerous proposals to quantify mixed-state entanglement \cite{PV07},
the logarithmic negativity \cite{VW02, Plenio05} turns out to be a particularly useful and easily
computable measure.
It is especially simple to evaluate for coupled harmonic oscillators \cite{AEPW02}
and in general for Gaussian states of continuous variable systems \cite{AI07}.
In particular, the logarithmic negativity was used to characterize tripartite entanglement
in the ground state of the one-dimensional harmonic chain \cite{MRPR09}. Furthermore,
results obtained for various quantum spin chains \cite{WMB09,WVB10} indicate universal features
in the entanglement of disjoint intervals at criticality.
These universal features have recently been understood via CFT methods \cite{CCT12,CCT13}; the corresponding analytical predictions have been compared to numerical simulations in various 1D critical
systems \cite{CTT13,Alba13,CABCL13} and a very good agreement was found. 
In two dimensions,
entanglement negativity has also been shown to detect topological order \cite{LV13,Castel13}.

In spite of this renewed interest, the behaviour of the entanglement negativity has 
not yet been
investigated out of equilibrium. Here we consider such a problem for a chain
of harmonic oscillators that is released from an initial state where the two sides of the chain
are kept at different temperatures. The system driven by this thermal gradient evolves,
for long times, into a nonequilibrium steady state (NESS) carrying a constant flux of energy.
In the gapless limit of the harmonic chain and for sufficiently low reservoir temperatures,
the NESS is believed to be described by
a nonequilibrium CFT with universal behaviour for e.g. the energy flow \cite{BD12,BD14} and thus
one expects that universal signatures may also appear in the entanglement negativity.

The choice of our nonequlibrium setup is further motivated by recent studies of a free-fermion
chain where the same type of NESS was shown to violate the area law for the mutual
information \cite{EZ14}. Similar logarithmic violations were found for the XY spin chain \cite{ABZ14}
which shows a marked contrast to the thermal-state behaviour where a strict area law can be proven
to hold \cite{WVHC08}. Since the mutual information measures only the total (classical + quantum)
correlations between subsystems, it is natural to ask whether such a singular behaviour
would be present for the steady-state entanglement negativity. The NESS of the harmonic chain is an
ideal candidate to attack this question since, in contrast to free fermions or spin chains, the logarithmic
negativity can easily be extracted from the covariance matrix \cite{AEPW02}.

Here we focus on the entanglement between adjacent segments of the chain and in a
low-temperature limit of the NESS.
Our main result is to show that the logarithmic negativity is, to a very good approximation,
a sum of two contributions from left- and right-moving normal-mode excitations emitted from the reservoirs.
They both carry one-half of the corresponding thermal-state entanglement that can be found from
a simple generalization of the ground-state CFT calculations \cite{CCT12,CCT13}.
Hence the steady-state entanglement of the harmonic chain obeys a strict area law.

In the next section we define the model and describe the initial and time evolved states, which is
followed by a discussion of the steady state properties for the infinite chain in Sec.~3.
Next, we present the covariance matrix
formalism in Sec.~4 that is used to obtain the logarithmic negativity. 
The method is then applied to calculate the steady-state entanglement in 
Sec.~5 whereas the full time evolution starting from
the initial state is presented in Sec.~6. The results are discussed in Sec.~7 
and some details of the
calculations are presented in two Appendices.

\section{Model and setting}

The Hamiltonian of the harmonic chain of length $N$ (with units $m=\hbar=1$) is given by
\eq{
H=\frac{1}{2}\sum_{n=0}^{N} \left[ p_n^2 + K ({x}_{n+1} - {x}_{n})^2 + \omega_0^2 x_n^2 \right],
\label{ham}}
where $x_n$ and $p_n$ denote the position and momentum operators of the $n$-th oscillator
with canonical commutation relations $\left[ x_m,p_n\right]=i \delta_{m,n}$ and
$\left[ x_m,x_n\right]=\left[ p_m,p_n\right]=0$. The parameters $K$ and $\omega_0^2$ set the strength
of the nearest neighbour coupling and the external harmonic confining potential at each site, respectively.
On the chain ends we impose Dirichlet boundary conditions (fixed walls)
which imply $x_{0} = x_{N+1}=0$ and $p_0=p_{N+1}=0$.

\begin{figure}[thb]
\center
\includegraphics[width=0.7\columnwidth]{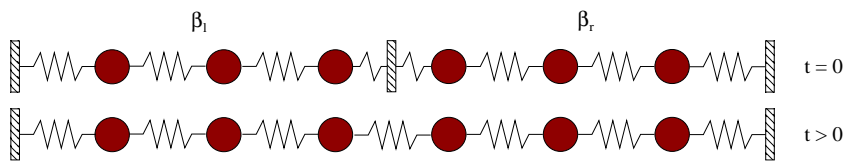}
\caption{Geometry of the oscillator chain before and after the quench.}
\label{fig:geom}
\end{figure}

The initial state is given by a simple product of Gibbs states
\eq{
\rho(0) = \frac{1}{Z_l} \ee^{-\beta_l H_l} \otimes
\frac{1}{Z_r} \ee^{-\beta_r H_r} \, ,
}
with inverse temperatures $\beta_l$ and $\beta_r$ for the left and right half-chains, respectively.
The Hamiltonian $H_l$ ($H_r$) is defined by a similar expression as in Eq.~(\ref{ham}) with the limits
of the sum given by $0,N/2$ ($N/2,N$) and Dirichlet boundary conditions imposed at sites
$0$ and $N/2+1$ ($N/2$ and $N+1$). 
The initially disconnected halves are joined together at time $t=0$ and the state evolves unitarily
$\rho(t) = \ee^{-i H t} \rho(0) \ee^{i H t} $ under the action of
Hamiltonian (\ref{ham}). The change in the geometry of the chain is depicted 
in Fig.~\ref{fig:geom}.

%%%%%%%%%%%%%%%%%%%%%%%%%%%%%%%%%%%%%%%%%%%%%%%%%%%%%%%%%%%
%
%
%%%%%%%%%%%%%%%%%%%%%%%%%%%%%%%%%%%%%%%%%%%%%%%%%%%%%%%%%%%

Since both the initial state as well as the time evolution operator is Gaussian, $\rho(t)$ can be
fully characterized by its $2N \times 2N$ covariance matrix $\Cov(t)$.
This can be written in a block-matrix form, composed of symmetric 
$2 \times 2$ matrices
\eq{
\Cov_{m,n}(t)= \langle R_m(t) R^T_n(t) + R_n(t) R^{T}_m(t)\rangle,
\quad R_n(t)= \twovec{x_n(t)}{p_n(t)},
%\twomat{\langle 2 x_m(t) x_n(t) \rangle}{\langle x_m(t) p_n(t)+p_n(t)x_m(t) \rangle}
%{\langle p_m(t) x_n(t)+x_n(t)p_m(t) \rangle}{\langle 2 p_m(t) p_n(t) \rangle}
\label{cov}}
%
%where $R_{2n-1}(t)=x_n(t)$ and $R_{2n}(t)=p_n(t)$ are the position and momentum operators
where $R_{n}(t)$ is a vector of the position and momentum operators
in the Heisenberg picture at site $n$, and
$\langle \mathcal{O}(t) \rangle =\Tr \left[ \rho(0) \mathcal{O}(t) \right]$ denotes 
the average of observable $\mathcal{O}(t)$ taken with the initial-state density operator.

To obtain the time evolution of $R_n(t)$, we first
introduce new operators by the canonical transformation
\eq{
x_k = \sum_{n=1}^{N} \phi_k(n) x_n, \quad
\phi_k(n) = \sqrt{\frac{2}{N+1}} \sin \frac{\pi k n}{N+1}, \quad
k = 1, \dots, N,
\label{phik}}
and similarly for $p_k$. Note, that $\phi_k(n)$ are just the eigenvectors of the dynamical matrix
in the Hamiltonian, satisfying Dirichlet boundary conditions $\phi_k(0)=\phi_k(N+1)=0$.
The Hamiltonian is then transformed into
\eq{
H = \frac{1}{2} \sum_{k=1}^{N} \left( p_k^2 + \omega_k^2 x_k^2 \right),
\label{hamk}}
with a dispersion relation
\eq{
\omega_k= \sqrt{\omega_0^2 + 4 K\sin^2(q_k/2)}, \qquad
q_k = \frac{\pi k}{N+1} .
\label{omk}}
In the following we will choose $K=1$, which sets the maximal velocity of the normal-mode
excitations to unity.

The Heisenberg equations of motion are given by $\dot x_k(t) = p_k(t)$ and
$\dot p_k(t) = -\omega_k^2 x_k(t)$, which yields $R(t)=S(t)R(0)$ with a symplectic
matrix $S(t)$ of block form
\eq{
%\fl
%\twomat{S_{2m{-}1, 2n{-}1}}{ S_{2m{-}1,2n}}{S_{2m,2n{-}1}}{S_{2m,2n}}
S_{m,n} (t)
=\sum_{k=1}^{N} \phi_k^*(m) \phi_k(n)
\twomat{\cos \omega_k t}{\omega_k^{-1} \sin \omega_k t }{-\omega_k\sin \omega_k t }{\cos \omega_k t}.
\label{st}}
The time-evolution of the covariance matrix then reads
\eq{
\Cov (t)=S(t)\Cov(0)S(t)^T, \qquad
\Cov(0) = \twomat{\Cov^l}{0}{0}{\Cov^r},
\label{ct}}
where $C(0)$ has a block-diagonal form, composed of $N \times N$ covariance matrices
of Gibbs states on the two disconnected half-chains. Their matrix elements are given by 
\eq{
%\fl
%\twomat{\Cov^{\alpha}_{2m{-}1, 2n{-}1}}{ \Cov^{\alpha}_{2m{-}1,2n}}
%{\Cov^{\alpha}_{2m,2n{-}1}}{\Cov^{\alpha}_{2m,2n}} 
C^{\alpha}_{m,n}
= \sum_{k=1}^{N/2} \phi^*_{\alpha,k}(m) \phi_{\alpha,k}(n)
\twomat{\omega_{\alpha, k}^{-1}}{0}{0}{\omega_{\alpha,k}}
\coth \frac{\beta_{\alpha} \omega_{\alpha,k} }{2},
\label{c0}}
with $\alpha=l,r$ and $1 \le m,n \le N/2$. Note, that $\phi_{\alpha,k}$ and $\omega^2_{\alpha,k}$
are the eigenvectors and eigenvalues of the dynamical matrix of the half-chains and are obtained
by exchanging $N \to N/2$ in Eqs.~(\ref{phik}) and (\ref{omk}), respectively.

\section{The steady state
\label{sec:ness}}

Our primary goal is to calculate the entanglement in the steady state, i.e., in the
asymptotic limit $t\to\infty$ of the time evolution. For this limit to be well defined,
one should set $1 \ll t \ll N$ in order to avoid reflections of the induced wavefront from
the fixed ends of the chain. This can be achieved by working directly in the
thermodynamic limit $N\to\infty$. The eigenvectors of the dynamical
matrix are then Fourier modes, $\phi_q(n)\sim \ee^{-iqn}$, and the corresponding limit
for the elements of the symplectic matrix in Eq.~(\ref{st}) is given by
\eq{
S_{m,n} (t) =
\int_{-\pi}^{\pi} \frac{\dd q}{2\pi} \ee^{iq(m-n)}
\twomat{\cos \omega_q t}{\omega_q^{-1} \sin \omega_q t }
{-\omega_q\sin \omega_q t }{\cos \omega_q t},
\label{sti}}
where $m,n \in \mathbb{Z}$ and the dispersion $\omega_q$ has the same form as in Eq.~(\ref{omk}), but
with continuous quasi-momenta $q \in (-\pi,\pi)$.

The time evolution on such an infinite chain was considered for the case of \emph{classical}
harmonic oscillators before, and the existence of a steady state was shown \cite{SL77,BPT83}. Note, however,
that the time evolution operator $S(t)$ is exactly the same for quantum oscillators and the only
difference between the two problems is the form of the initial covariance matrix $C(0)$.
Moreover, it was shown in Ref.~\cite{BPT83} that, for a large class of initial conditions, the covariance
matrix converges \emph{locally} under the time evolution $S(t)$ in the limit $t \to \infty$.
This is true, in particular, if $C(0)$ is translationally invariant asymptotically far away from the cut
on both left- and right-hand sides, which is satisfied by the Gibbs-state covariances in Eq. (\ref{c0})
in the limit $N\to\infty$ and $|m|,|n| \gg 1$.

The formal derivation of $C(t)$ in the limit $t\to\infty$ was given in 
Ref.~\cite{BPT83}. However, there is a small mistake in the calculation and thus we reiterate the main steps with correct formulas
in \ref{app:sscov}. In turn, the asymptotic limit of the covariance matrix is obtained as
\eq{
\lim_{t\to\infty} \Cov_{m,n} (t) =  \int_{-\pi}^{\pi}  \frac{\dd q}{2 \pi} \ee^{i q (m-n)}
\left( C^{+}_q + i \, {\rm sgn} (q) \, C^{-}_q \right),
\label{cinf}}
with $2 \times 2$ matrices
\begin{eqnarray}
C^{+}_q = \frac{1}{2} \left[
\coth \left( \frac{\beta_r \omega_q }{2}\right) +
\coth \left( \frac{\beta_l \omega_q}{2}\right) \right]
\twomat{ \omega_q^{-1}}{0 }{ 0 }{\omega_q} ,\label{cpq}\\
C^{-}_q = \frac{1}{2} \left[
\coth \left( \frac{\beta_r \omega_q }{2}\right) -
\coth \left( \frac{\beta_l \omega_q}{2}\right) \right]
\twomat{ 0 }{ -1 }{1 }{ 0 }.
\label{cmq}
\end{eqnarray}
Note, that the limit in (\ref{cinf}) holds by fixed indices $m,n$ and gives a steady-state
covariance matrix which is locally translational invariant.

In case $\beta_l \ne \beta_r$, the steady state supports a nonzero energy current
which is encoded in the offdiagonal matrix elements of Eq.~(\ref{cmq}).
The nature of the NESS is however best understood by rather considering the correlation
functions of the bosonic annihilation $a_m$ and creation $a_m^\dag$ operators, defined
as the Fourier transforms of the modes
\eq{
a_q = \sqrt{\frac{\omega_q}{2}}\left( x_q + \frac{i}{\omega_q} p_q \right), \qquad
a_q^{\dag} = \sqrt{\frac{\omega_q}{2}}\left( x_{-q} - \frac{i}{\omega_q} p_{-q} \right),
\label{aq}}
which bring the Hamiltonian (\ref{hamk}) into diagonal form. The asymptotic form of the
bosonic correlation functions is obtained as
\eq{
\lim_{t\to\infty} \Tr (\rho(t) \, a^{\dag}_{m} a_{n}) = 
\int_{-\pi}^{\pi}  \frac{\dd q}{2 \pi} \ee^{i q (m-n)} \, n_q \, ,
\label{ainf}}
where the bosonic mode-occupation number is given by
\eq{
n_q =
\cases{\frac{1}{\ee^{\beta_r \omega_q}-1} & $q \in \left( -\pi, 0 \right)$, \\
\frac{1}{\ee^{\beta_{l} \omega_q}-1} & $q \in \left( 0, \pi \right)$. }
\label{nq}}

The form of $n_q$ has a simple physical interpretation. Namely, all the normal modes
with positive (negative) group velocities $\frac{\dd \omega_q}{\dd q}$ originate from the
left (right) reservoir and thus their occupation numbers are given by the corresponding Bose-Einstein
statistics with inverse temperature $\beta_l$ ($\beta_r$). Note, that this behaviour is completely
analogous to the one found for free-fermion related models \cite{AH00,Ogata02,AP03,DVBD14,CK14,CM14},
and also agrees with recent results obtained within CFT \cite{BD12,BD14}.

\subsection{GGE form of the steady state}

The steady state given by the bosonic mode-occupation numbers (\ref{nq}) does not correspond to
a Gibbs ensemble, unless $\beta_l = \beta_r$. However, taking into account all the (infinite set of)
local conservation laws for the harmonic chain, one can express the NESS density matrix as a
generalized Gibbs ensemble (GGE) \cite{BPT83}
\eq{
\lim_{t\to\infty}\rho (t) = \frac{1}{Z} e^{- H_{{\rm eff}}}, \qquad
H_{{\rm eff}}=\sum_{n=0}^{\infty} (\mu_n^{+} Q^{+}_n + \mu^-_n Q^{-}_n) \, ,
\label{gge}}
%
%where $\beta=(\beta_\ell + \beta_r)/2$
where the integrals of motion can be constructed in the
$N \to \infty$ limit of a periodic chain and read \cite{BPT83}
\begin{eqnarray}
Q^+_n = \frac{1}{2}\sum_{m=-\infty}^{\infty} \left[ p_m p_{m+n} {+}
(x_m {-}x_{m+1})(x_{m+n}{-}x_{m+n+1}) {+}  \omega^2_0x_mx_{m+n}\right],\\
Q^-_n = \frac{1}{4}\sum_{m=-\infty}^{\infty}
%(x_m p_{m+n} - p_m x_{m+n} - x_{m+n}p_{m} + p_{m+n}x_m ).
(p_m x_{m+n} - x_m p_{m+n} - p_{m+n}x_m + x_{m+n}p_{m} ).
\label{qpm}
\end{eqnarray}
%
%\eq{
%Q^+_n = \sum_{q>0} \omega_q \left( a_q a_q^{\dag} + a_{-q}^{\dag}a_{-q}\right) \cos nq
%}
%\eq{
%Q^-_n = \sum_{q>0} \left( a_q a_q^{\dag} - a_{-q}^{\dag}a_{-q}\right) \sin nq
%}
%
%\eq{
%\mu_n^- = \frac{\beta_l-\beta_r}{2} \int_{0}^{\pi} \frac{\dd q}{\pi} \omega_q \sin nq
%}
The corresponding sets of Lagrange multipliers $\mu_n^{\pm}$ can be determined by taking the
Fourier transform of $H_{\mathrm{eff}}$, rewriting it in terms of the bosonic operators $a_q$ and $a_q^\dag$,
and requiring that the resulting expression reproduces the mode-occupations in Eq.~(\ref{nq}).
This leads to the equations
\begin{eqnarray}
\sum_{n=0}^{\infty} 
\left( \mu_n^+ \omega_q \cos nq +  \mu_n^- \sin nq \right) = \omega_q \beta_l \, ,\\
\sum_{n=0}^{\infty} 
\left( \mu_n^+ \omega_q \cos nq -  \mu_n^- \sin nq \right) = \omega_q \beta_r ,
\label{mun1}
\end{eqnarray}
which can be solved and yield the Lagrange multipliers
\eq{
\mu_{n}^{+} = \frac{\beta_l+\beta_r}{2} \delta_{n,0} \, , \qquad
\mu_{n}^{-} = 
%(-1)^{n+1}\frac{4}{\pi}\frac{\beta_{\ell}-\beta_r}{\beta_{\ell}+\beta_r}\frac{2n}{4n^2-1}
( \beta_l-\beta_r ) (-1)^{n} \frac{4}{\pi} \frac{n}{4n^2-1} .
\label{mun2}}
Note, that the only reflection-symmetric conserved charge appearing in the GGE
is $Q_0^+ = H$ the original Hamiltonian itself and the corresponding multiplier is simply the average
inverse temperature. On the other hand, all the $Q_n^-$ charges have nonvanishing multipliers,
decaying slowly $\mu_n^- \sim 1/n$ for $n \gg 1$ and alternating in sign.
Therefore, the operator $H_{\mathrm{eff}}$ has long-range couplings.
Interestingly, the steady state of a free-fermion chain, starting from the same initial
condition, has a completely analogous GGE description \cite{EZ14, Ogata02}, the only difference
being that multipliers with odd indices vanish identically there, and thus one has no
sign alternation in $H_{\mathrm{eff}}$.

To conclude this section, we remark that the non-local structure of the GGE is a direct consequence
of the thermodynamic limit. Indeed, if boundary conditions are retained at the ends of the chain,
one expects that the currents are suppressed for large times due to reflections, $\mu_n^- =0$ for all $n$, and thus the GGE becomes local.
This was recently proven for a free-fermion field theory \cite{CK14} and we believe that the same holds also in the continuum limit of the oscillator chain.

\section{Partial transpose and logarithmic negativity
\label{sec:ln}}

Our aim is to characterize the amount of entanglement in the time-evolved state $\rho(t)$ of the harmonic chain.
In the following, we focus on a particular measure of entanglement, the logarithmic negativity \cite{VW02}.
In the most general case, one is interested in a tripartite setting where entanglement is to be measured
between subsystems $A_1$ and $A_2$, with $A=A_1 \cup A_2$ and $B$ denoting the rest of the chain.
The logarithmic negativity is then defined through the partial transpose of the reduced density matrix
$\rho_A(t)=\Tr_B \rho(t)$ as
\eq{
\mathcal{E} = \ln \Tr \left| \rho_A^{T_2}(t) \right|,
\label{lnrho}}
where the superscript $T_2$ indicates a partial transposition with respect to the indices in subsystem $A_2$.
The logarithmic negativity thus detects only the negative eigenvalues 
of $\rho_A^{T_2}(t)$.
In particular, if there is no entanglement between $A_1$ and $A_2$, then all the eigenvalues of $\rho_A^{T_2}(t)$
are positive \cite{Peres96} and $\mathcal{E}$ vanishes due to normalization.

Instead of working with density matrices, the logarithmic negativity of the harmonic chain is easier
to obtain using the covariance matrix formalism \cite{AEPW02}. Indeed, the eigenvalues of $\rho_A(t)$ are
related to the symplectic eigenvalues of the reduced covariance matrix $\Cov_A(t)$. Moreover, the
partial transpose can also be implemented on the level of the covariances, with $\Cov_A^{T_2}(t)$
denoting the matrix where the signs of all the momenta $p_n$ with $n \in A_2$ are reversed.
In turn, the logarithmic negativity is obtained as \cite{AEPW02}
\eq{
\mathcal{E} = - \sum_{j=1}^{|A|} \ln \min (\nu_j,1)
\label{lnnu}}
from the symplectic eigenvalues $\nu_j$ of $\Cov_A^{T_2}(t)$ and $|A|$ denotes the number of sites in $A$.
Note, that only the eigenvalues $\nu_j < 1$ contribute.

The symplectic spectrum of $\Cov_A^{T_2}(t)$ can be obtained through ordinary diagonalization,
by multiplying with the symplectic matrix
\eq{
\Sigma_A = \bigoplus_{j \in A}  \twomat{0}{1}{-1}{0},
\label{sigma}}
which leads to the spectrum
\eq{
\mathrm{Spect} \left( \Sigma_A \Cov_A^{T_2}(t) \right) = \left\{ \pm i\nu_1, \dots, \pm i\nu_{|A|} \right\}.
\label{spect}}

\section{Steady-state entanglement
\label{sec:lnness}}

We are interested in the entanglement of two neighbouring subsets of oscillators
$A_1$ and $A_2$, each of size $\ell$.
Before presenting results for the logarithmic negativity $\mathcal{E}$, one should point out
a subtlety of the numerical treatment. In all the following calculations, we are interested
in the gapless limit of the harmonic chain, $\omega_0 \to 0$, where the Hamiltonian has
an underlying CFT with central charge $c=1$. Note, however, that both expressions (\ref{cpq})
and (\ref{cmq}) diverge for $q \to 0$ in this limit due to the zero-mode. Nevertheless,
we observe numerically that $\mathcal{E}$ is insensitive to the zero-mode contribution
and converges as $\omega_0 \to 0$. Interestingly, this is in contrast with the behaviour
of the entropy and mutual information, which both involve a divergent zero-mode contribution.
To ensure that we always remain in the gapless regime,
all the calculations are carried out by setting $\omega_0 = 0.005/\ell$, i.e., reducing the gap
with the size of the subsystems.

\subsection{Equal temperatures}

We first consider the simplest case $\beta_l=\beta_r=\beta$. The
local perturbation in the center, due to the change in the boundary condition,
is propagated away and the NESS converges locally to the Gibbs equilibrium state
$\rho \sim \exp(-\beta H)$.
The thermal-state entanglement in both spin models \cite{TKGB07} and oscillator
chains \cite{AW08,Anders08,FCGA08,CFGA08}  has been studied previously
with a focus on the critical temperature above which the state becomes separable.

Here, instead, we consider the scaling of the entanglement negativity of two adjacent intervals
in the low-temperature regime, $\beta \gg 1$, as a function of $\ell$ and $\beta$.
It turns out that CFT methods, applied recently to calculate the ground-state
entanglement in the oscillator chain \cite{CCT13}, can easily be generalized to finite temperatures
in this case. In particular, for two adjacent intervals of lengths $\ell_1$ and $\ell_2$
embedded in an infinite chain, the ground-state logarithmic negativity can be expressed via
three-point functions of twist fields on the complex plane and yields \cite{CCT13}
\eq{
\mathcal{E}
%_\mathrm{gs}
=\frac{c}{4} \ln \frac{\ell_1 \ell_2}{\ell_1 + \ell_2} + \mathrm{const.}
\label{lnl12}}
where $c$ is the central charge of the CFT. For finite temperatures, the CFT is defined on an
infinite cylinder of circumference $\beta$ which, however, can be mapped into the plane by
a simple conformal transformation. The overall effect of the mapping is to replace the
lengths $\ell_i \to \beta/\pi \sinh ( \ell_i \pi/ \beta )$. For intervals of equal lengths,
$\ell_1 = \ell_2 = \ell$, this leads to
\eq{
\mathcal{E} = \frac{1}{4} \ln \frac{\beta}{\pi} \tanh \frac{\ell \pi}{\beta} + \mathrm{const.}
\label{lnbeta}}
where we have set $c=1$ corresponding to the harmonic chain.

\begin{figure}[thb]
\center
\psfrag{E}[][][.7]{$\mathcal{E}$}
\psfrag{L}[][][.7]{$\ell$}
\psfrag{x}[][][.7]{$\beta/\pi \tanh(\ell \pi / \beta)$}
\includegraphics[width=0.49\columnwidth]{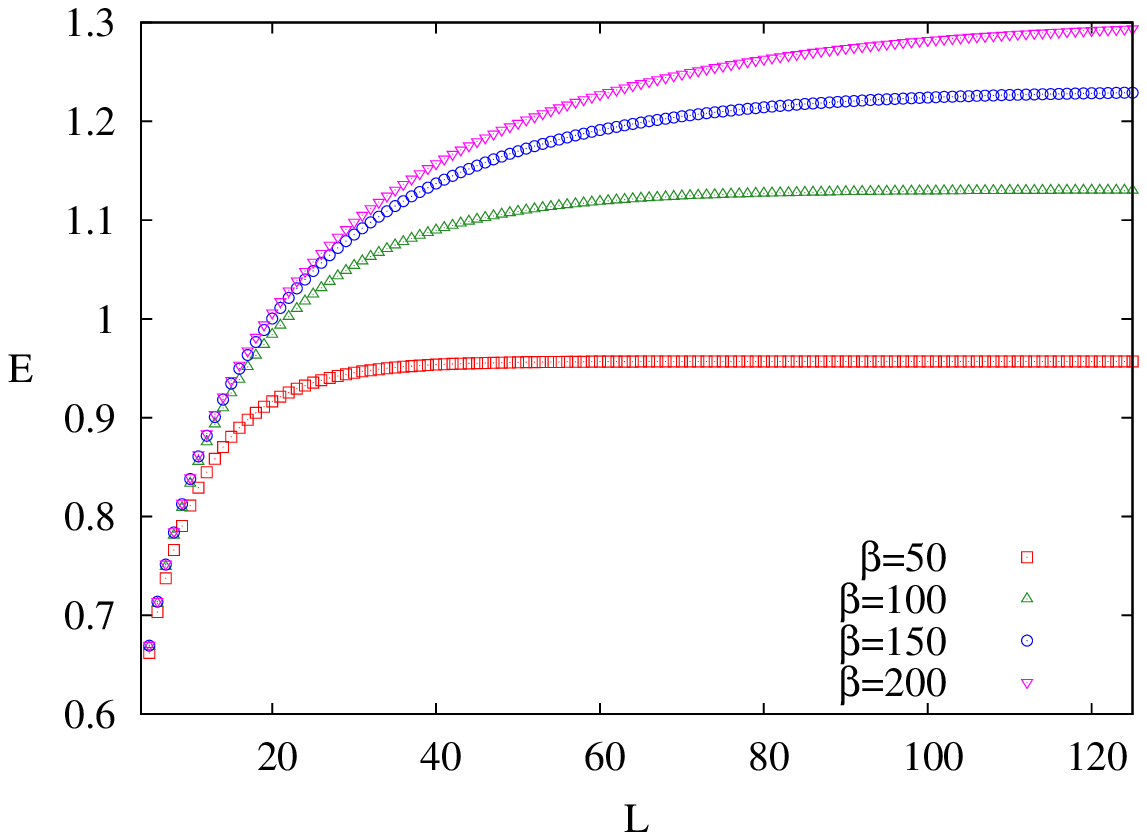}
\includegraphics[width=0.49\columnwidth]{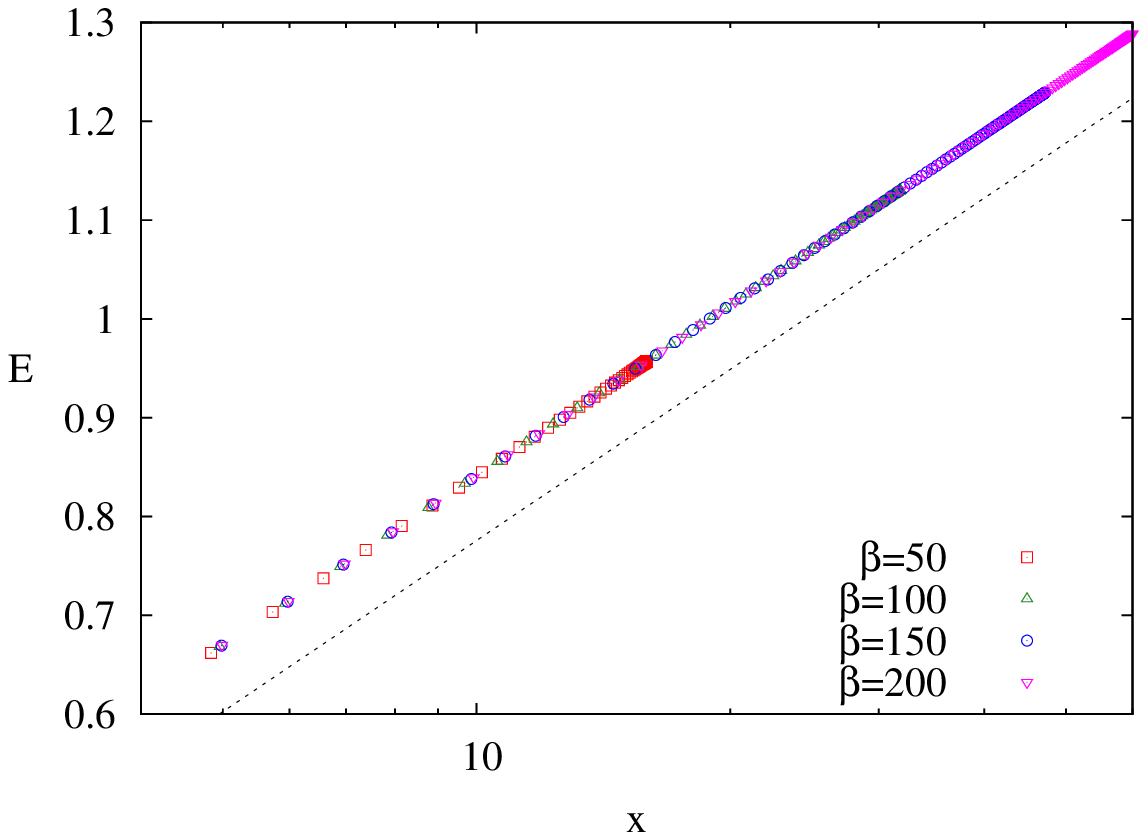}
\caption{Left: Thermal-state logarithmic negativity for two adjacent intervals of size $\ell$
in an infinite chain for various $\beta$.
Right: Scaled data according to CFT prediction.
% on a semi-logarithmic scale.
The dotted line has slope $1/4$.}
\label{fig:lnbeta}
\end{figure}

The CFT formula (\ref{lnbeta}) has the correct limiting behaviour
$\mathcal{E}\sim 1/4 \ln \ell$ for $\ell \ll \beta$ whereas in the opposite limit 
of large segment sizes, $\ell \gg \beta$, it predicts the saturation value
$\mathcal{E}\sim 1/4 \ln \beta$. This is indeed what we observe from the numerical data,
obtained by the method in Section \ref{sec:ln}, and shown on the left of 
Fig.~\ref{fig:lnbeta}.
When scaled according to the CFT variable, as shown on the right of 
Fig.~\ref{fig:lnbeta}, we observe an excellent data collapse and agreement 
with the prediction of Eq.~(\ref{lnbeta}).

%%%%%%%%%%%%%%%%%%%%%%%%%%%%%%%%%%%%%%%%%%%%%%%%%%%%%%%%%%%
%
%
%%%%%%%%%%%%%%%%%%%%%%%%%%%%%%%%%%%%%%%%%%%%%%%%%%%%%%%%%%%

\subsection{Unequal temperatures}

Turning to the nonequilibrium case, we now present some simple heuristic arguments how the
steady-state entanglement can be related to the equilibrium value in Eq.~(\ref{lnbeta}).
Following the results of Section \ref{sec:ness}, the NESS density operator can be written
in the form $\rho \sim \exp \left(-\beta_l H_+ -\beta_r H_-\right)$ with $H_{\pm}$ describing the evolution
of right- and left-moving particles, respectively. In the CFT context \cite{BD14}, they correspond to mutually
commuting chiral components of the Hamiltonian $H=H_+ + H_-$. In the path-integral representation of
$\rho$, the action thus decouples in the chiral fields $\phi_{+}$ and $\phi_{-}$ that live on infinite cylinders
of circumferences $\beta_l$ and $\beta_r$, respectively. Due to this separation, the partition functions involved
in the calculation of the logarithmic negativity \cite{CCT13} are also supposed to factorise 
into chiral components
and thus their contribution is additive. Finally, making the natural assumption that the entanglement
content of the chiral theories is half of that of the full CFT, one expects the relation
\eq{
\mathcal{E}(\beta_l,\beta_r) = \frac{\mathcal{E}(\beta_l) + \mathcal{E}(\beta_r)}{2} \, ,
\label{lnbetalr}}
with the steady-state entanglement $\mathcal{E}(\beta_l,\beta_r)$ being the average of the
thermal-state values (\ref{lnbeta}) corresponding to the left and right reservoirs.

\begin{figure}[thb]
\center
\psfrag{j}[][][.7]{$j$}
\psfrag{nj}[][][.7]{$\nu_j(\beta_l,\beta_r)$}
\psfrag{dnj}[][][.6]{$\nu_j(\beta_l,\beta_r)-\sqrt{\nu_j(\beta_l) \nu_j(\beta_r)}$}
\includegraphics[width=0.6\columnwidth]{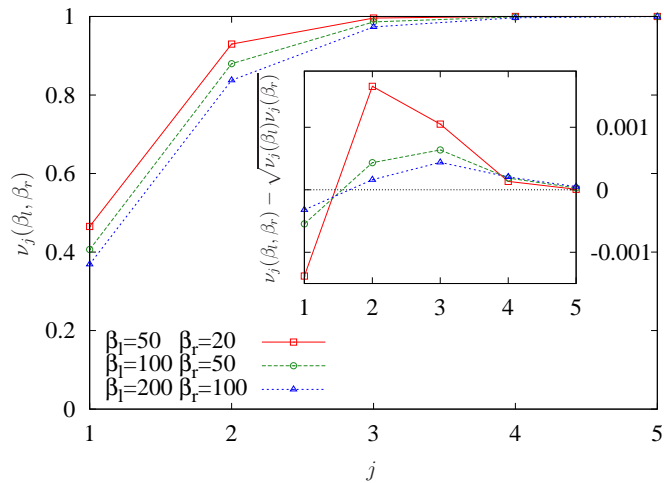}
\caption{The five smallest symplectic eigenvalues $\nu_j(\beta_l,\beta_r)<1$ of the
partial transposed steady-state covariance matrix
for two adjacent intervals of size $\ell=100$ and various pairs of $\beta_l$ and $\beta_r$.
The inset shows the deviation from the geometric mean of Gibbs-state symplectic
eigenvalues $\nu_j(\beta_l)$ and $\nu_j(\beta_r)$.}
\label{fig:spect}
\end{figure}

Our numerical calculations confirm the validity of Eq.~(\ref{lnbetalr}) to an 
extremely good precision.
The tiny deviations from the equality are supposed to be a consequence of the 
zero-mode which has been
neglected in the above CFT reasoning. In fact, the presence of the zero-mode couples the two chiral branches
$H_{\pm}$ and thus the factorization of the NESS density matrix is not perfect. However, the effect of this coupling
seems to be rather small for the range of temperatures we have considered.
This is illustrated in Fig.~\ref{fig:spect} on the level of the symplectic eigenvalues $\nu_j(\beta_l,\beta_r)<1$
of the partial transposed covariance matrix which contribute to $\mathcal{E}(\beta_l,\beta_r)$, see Eq.~(\ref{lnnu}).
The small deviations from the geometric mean $\sqrt{\nu_j(\beta_l) \nu_j(\beta_r)}$
of the Gibbs-state symplectic eigenvalues are shown on the inset. The deviations seem to increase
with increasing temperatures and the relation is expected to break down approaching the
critical temperature where the entanglement vanishes \cite{AW08,CFGA08}.

%%%%%%%%%%%%%%%%%%%%%%%%%%%%%%%%%%%%%%%%%%%%%%%%%%%%%%%%%%%
%
%
%%%%%%%%%%%%%%%%%%%%%%%%%%%%%%%%%%%%%%%%%%%%%%%%%%%%%%%%%%%

\section{Time evolution of entanglement}

We now consider the complete time evolution of $\mathcal{E}$ after the wall
separating the two sides of the chain is removed, see Fig.~\ref{fig:geom}.
The subsystems $A_1$ and $A_2$ are chosen to be adjacent intervals with their
common boundary located at the initial cut. Clearly, at $t=0$ one has
$\mathcal{E}=0$ and the entanglement has to increase to reach its steady-state
value, discussed in the previous section.

\subsection{Local quench at zero temperature}

The special case, when both sides of the chain are prepared in their respective ground states
will be treated first. In fact, this is the same situation, also known as a local quench, which
was studied before for free-fermion chains \cite{EP07,EKPP08}, as well as in the context of CFT
\cite{CC07,SD11}, in various geometries and with a focus on the entanglement entropy.

In some simple bipartite settings, $B = \emptyset$, the result can directly be inferred from
previous CFT calculations.
The simplest choice is to consider the evolution of $\mathcal{E}$ for two halves of an
infinite system, $A_1=\left(-\infty,0\right]$ and $A_2=\left[1,\infty\right)$. Since the state
$\rho(t)$ is pure, the logarithmic negativity is just the R\'enyi entropy with index $1/2$,
and inserting this into the CFT formula of Ref.~\cite{CC07} one obtains
\eq{
\mathcal{E} = \frac{1}{2} \ln t  + \mathrm{const.}
\label{lnlqh}}
Alternatively, one can follow the line of Ref.~\cite{CCT13} and work out the CFT representation
of the partial-transpose density matrix. The calculation is sketched in \ref{app:cft} and leads
to the same result.

\begin{figure}[htb]
\center
\psfrag{L=25}[][][.7]{$\ell = 25$}
\psfrag{L=50}[][][.7]{$\ell = 50$}
\psfrag{L=75}[][][.7]{$\ell = 75$}
\psfrag{L=100}[][][.7]{$\ell = 100$}
\psfrag{t}[][][.7]{$t$}
\psfrag{x}[][][.7]{$2\ell/\pi \sin(\pi t/2\ell)$}
\psfrag{E}[][][.7]{$\mathcal{E}$}
\includegraphics[width=0.49\columnwidth]{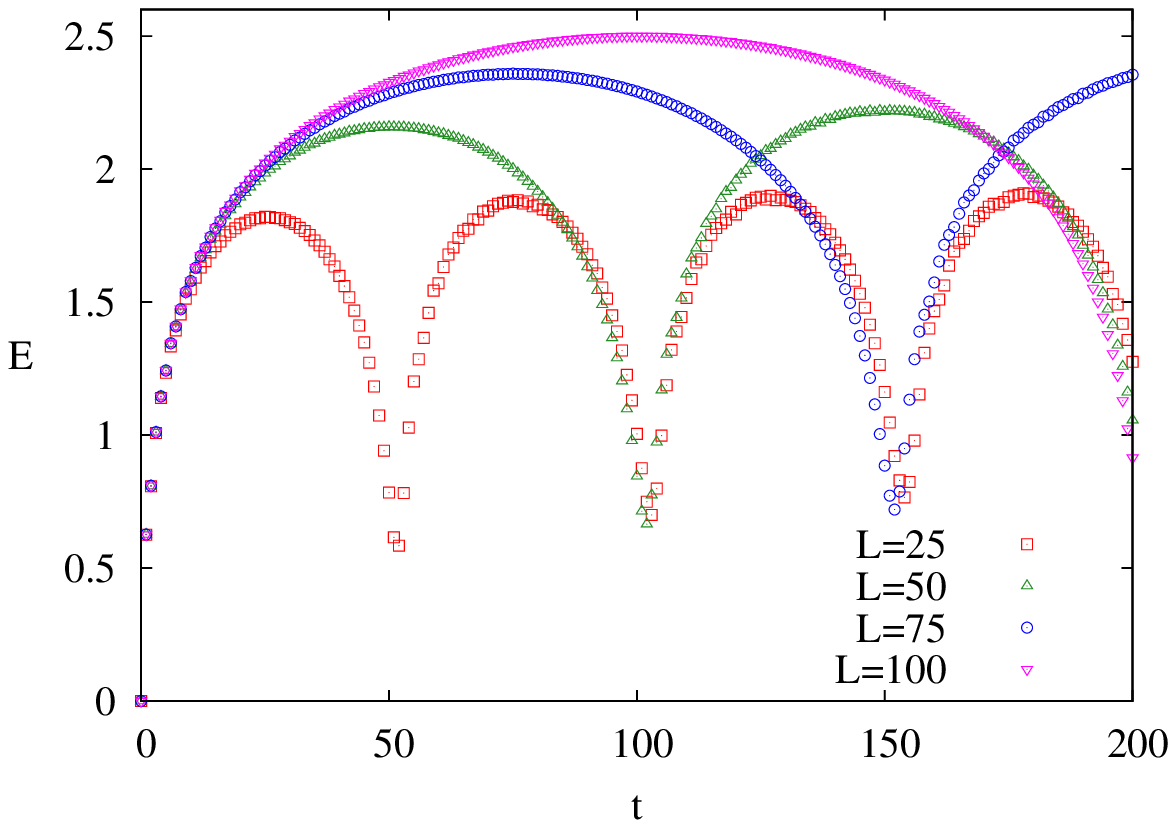}
\includegraphics[width=0.49\columnwidth]{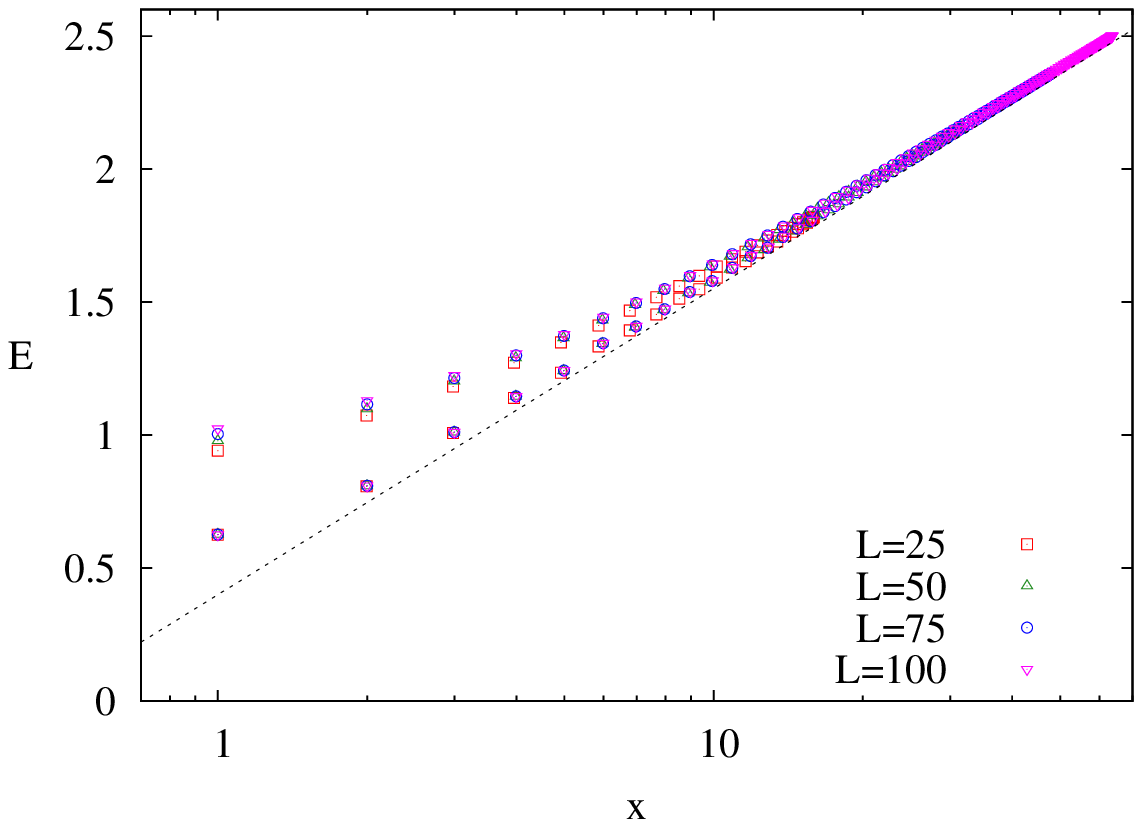}
\caption{Left: Time evolution of the half-chain logarithmic negativity after a local quench
in the finite geometry for various $\ell$. Right: Scaled data according to CFT prediction for
times $t < 2\ell$. The dotted line has slope 1/2.}
\label{fig:lnlqf}
\end{figure}

In order to test the result numerically, however, one has to choose a finite system of size $N=2\ell$,
with a bipartition $A_1=\left[1,\ell\right]$ and $A_2=\left[\ell+1,2\ell\right]$ with the initial cut
located between sites $\ell$ and $\ell+1$. The corresponding result for $\mathcal{E}$ can also be
found from a CFT calculation based on Ref.~\cite{SD11} and reads
\eq{
\mathcal{E}= \frac{1}{2} \ln \left| 2\ell/\pi \sin(\pi t/2\ell) \right| + \mathrm{const.}
\label{lnlqf}}
This can now be compared to numerical results obtained with the exact time-dependent
covariance matrix, Eqs.~(\ref{st}-\ref{c0}), following again the steps in 
section \ref{sec:ln},
with the result shown in Fig.~\ref{fig:lnlqf}. On the left, the data is plotted for various half-chain
lengths $\ell$ and shows a quasi-periodic structure with period $2\ell$, similar to the evolution of
entanglement entropy in the same geometry. This is due to reflections
of the front, induced by the quench, from the fixed ends of the chain. Note, however, the slight upwards
shift of the curve for $\ell=25$ which is supposedly due to lattice effects, caused by the slower normal-mode excitations with velocity
$\frac{\dd \omega_q}{\dd q} < 1$. Furthermore, there might be universal subleading contributions,
originating from a more careful CFT treatment and breaking the periodicity \cite{SD13}, that are,
however, difficult to identify in the numerics.
Nevertheless, when plotted against the CFT scaling variable in Eq.~(\ref{lnlqf}), the data shows a
very good collapse and is seen to reproduce the formula for large arguments,
as shown on the right of Fig.~\ref{fig:lnlqf}. 

%%%%%%%%%%%%%%%%%%%%%%%%%%%%%%%%%%%%%%%%%%%%%%%%%%%%%%%%%%%
%
%
%%%%%%%%%%%%%%%%%%%%%%%%%%%%%%%%%%%%%%%%%%%%%%%%%%%%%%%%%%%

Finally, one could consider $\mathcal{E}$ for the infinite chain in the tripartite setting
of section \ref{sec:lnness}. As discussed in \ref{app:cft}, the CFT treatment is rather involved,
requiring knowledge of higher order twist-field expectation values. Thus, we resort to the
numerical evaluation of $\mathcal{E}$. The ground-state covariance matrix of the semi-infinite
chain, with matrix elements given by the $N\to\infty$ limit of Eq.~(\ref{c0}), 
can be evaluated
explicitly for $\omega_0=0$ and yields \cite{CCT13}
\eq{
%\fl
C^{r}_{m,n} = C^r(m+n) - C^r(m-n), \quad
C^r(x) = \frac{1}{\pi} \twomat{\psi(1/2+x)}{0}{0}{\frac{4}{4x^2-1}}
\label{c0si}}
for the right-hand side of  the chain, $m,n \ge 1$, with $\psi(z)$ being the digamma function.
The left-hand side covariance matrix $C^l_{m,n}$ for $m,n \le 0$ is obtained by a reflection of
the indices $m \to 1-m$ and $n \to 1-n$ in Eq.~(\ref{c0si}).

The time evolved covariance matrix, Eq.~(\ref{ct}), requires to carry out the 
matrix product
with the symplectic evolution matrix $S(t)$ which, in principle, is infinitely large.
However, we can make use of the light-cone structure of the matrix elements, implying
that for $|m-n| \gg t$ the entries $S_{m,n}(t)$ are exponentially small. Thus, the sums
involved in $\Cov_A^{T_2}(t)$ can be truncated and evaluated numerically to a very high precision.

\begin{figure}[htb]
\center
\psfrag{L=25}[][][.7]{$\ell = 25$}
\psfrag{L=50}[][][.7]{$\ell = 50$}
\psfrag{L=75}[][][.7]{$\ell = 75$}
\psfrag{L=100}[][][.7]{$\ell = 100$}
\psfrag{t}[][][.7]{$t$}
\psfrag{x}[][][.7]{$t^{a} (\ell-t)^{b} (\ell+t)^{c} / \ell^{b+c}$}
\psfrag{E}[][][.7]{$\mathcal{E}$}
\includegraphics[width=0.49\columnwidth]{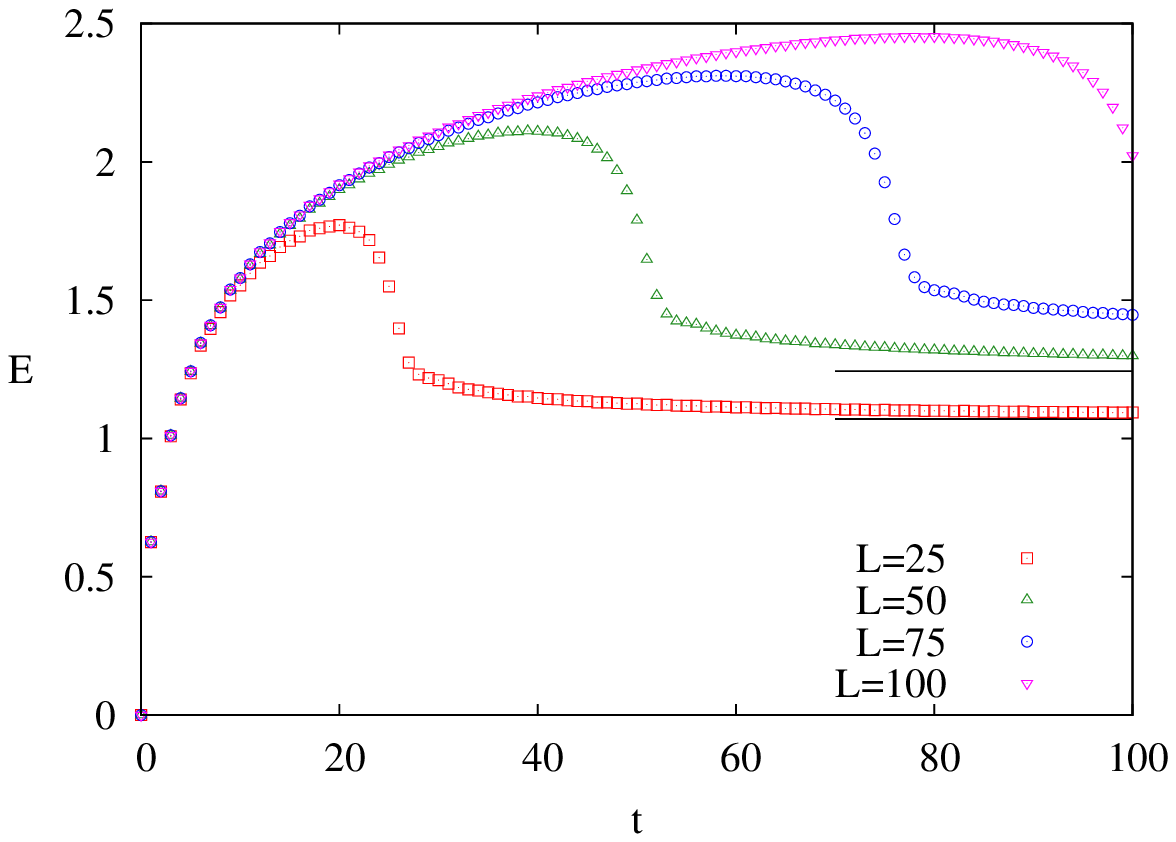}
\includegraphics[width=0.49\columnwidth]{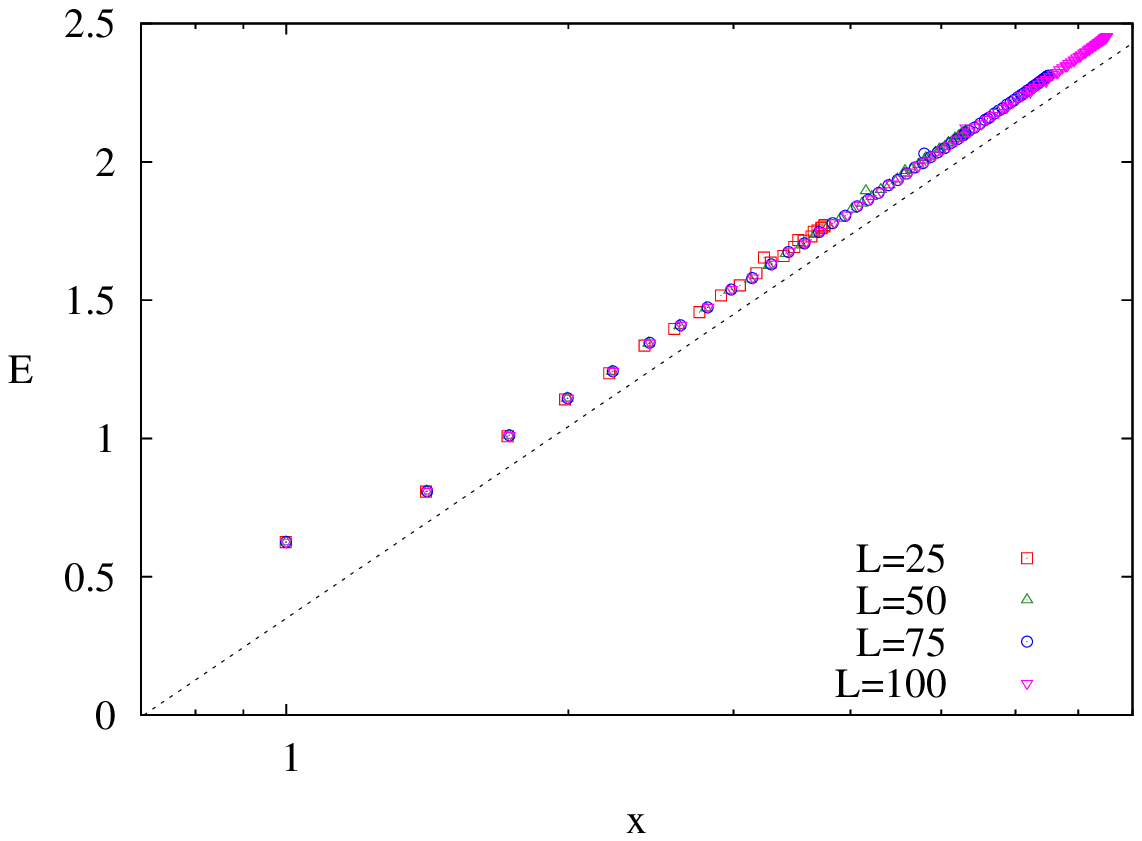}
\caption{Left: Time evolution of entanglement after a local quench in the infinite
geometry for various $\ell$. Horizontal lines indicate the ground-state values of
$\mathcal{E}$ for $\ell = 25,50$. Right: Scaled data according to Eq.~(\ref{lnlqi}) with
parameters $a=0.5$, $b\approx 0.15$, $c \approx 0.13$ and $d=-(b+c)$.}
\label{fig:lnlqi}
\end{figure}

The result for $\mathcal{E}$ is shown in Fig.~\ref{fig:lnlqi}. 
For times $t < \ell$, the logarithmic
negativity develops a plateau, followed by a sharp drop at $t\approx\ell$, where
the propagating front leaves the subsystem $A$. The data then converges slowly to the
ground-state value of $\mathcal{E}$, shown by horizontal lines on the left of 
Fig.~\ref{fig:lnlqi}.
The plateau region is again reminiscent of the behaviour of the entanglement entropy
in the corresponding geometry \cite{EP07,CC07}. We thus propose the ansatz
\eq{
\mathcal{E}= a \ln t + b \ln (\ell - t) + c \ln (\ell + t) + d \ln \ell + \mathrm{const.}
\label{lnlqi}}
where the coefficients are fitting parameters. In fact, some of them can be fixed by
requiring that in the limit $\ell \to \infty$ we recover the result (\ref{lnlqh}),
implying $a=1/2$ and $d=-(b+c)$. For the remaining two we obtain $b\approx 0.15$
and $c\approx 0.13$ from a fit to the data with $\ell=100$. The data is then
scaled together using these values in the right of 
Fig.~\ref{fig:lnlqi} and shows a nice collapse.

%%%%%%%%%%%%%%%%%%%%%%%%%%%%%%%%%%%%%%%%%%%%%%%%%%%%%%%%%%%
%
%
%%%%%%%%%%%%%%%%%%%%%%%%%%%%%%%%%%%%%%%%%%%%%%%%%%%%%%%%%%%

\subsection{Finite temperatures}

\begin{figure}[htb]
\center
\psfrag{t}[][][.7]{$t$}
\psfrag{E}[][][.7]{$\mathcal{E}$}
\includegraphics[width=0.49\columnwidth]{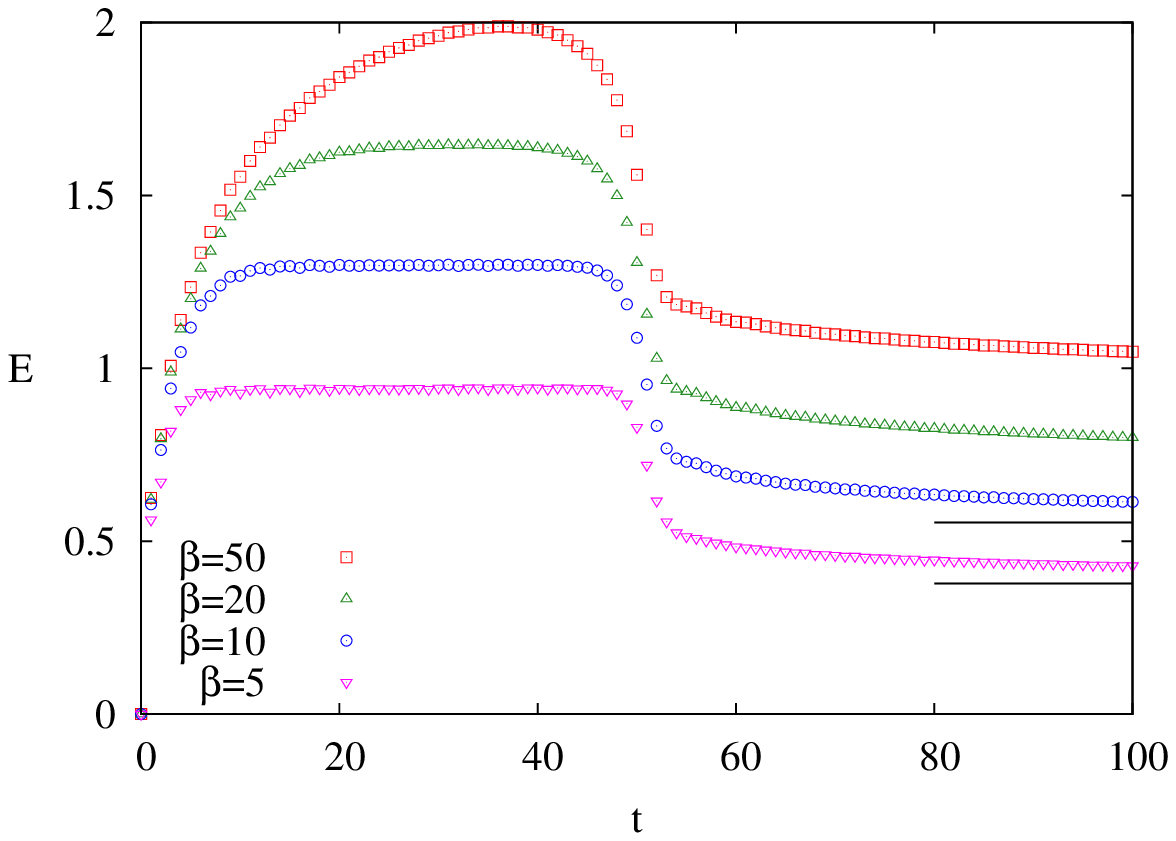}
\includegraphics[width=0.49\columnwidth]{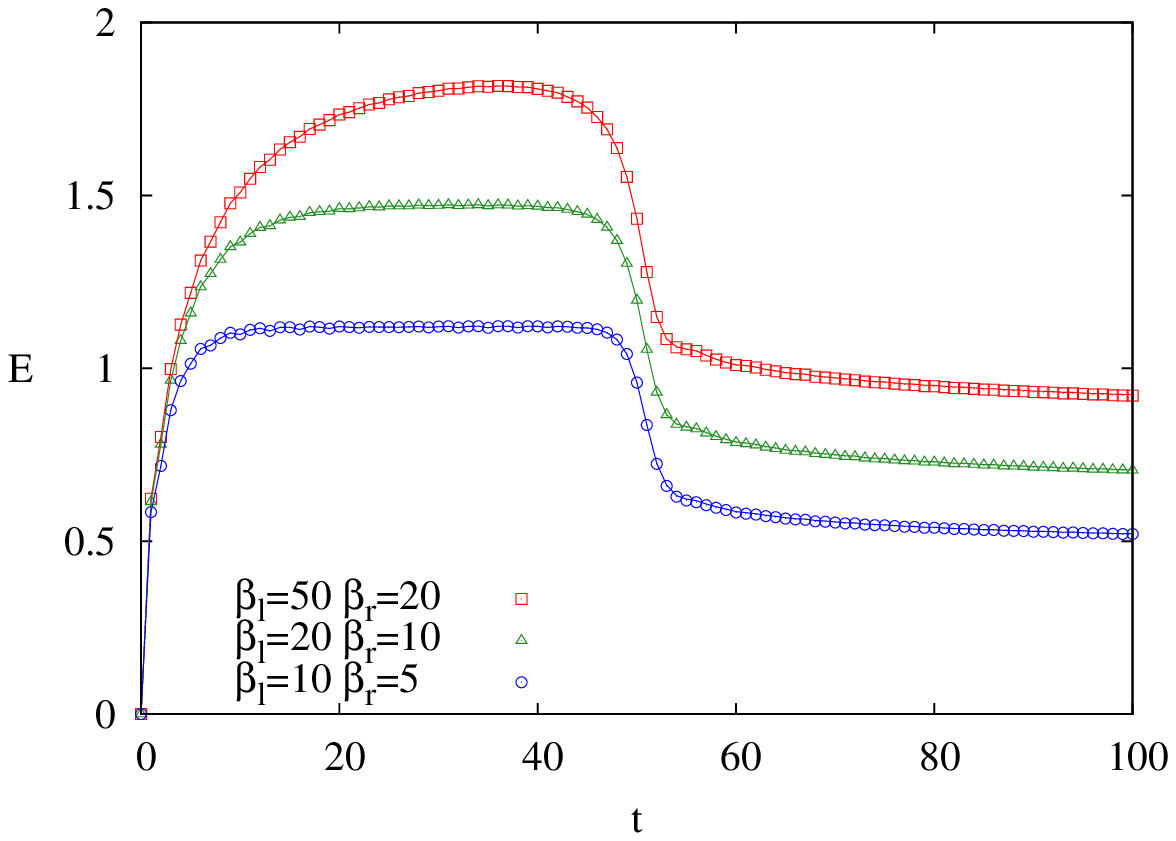}
\caption{Time evolution of entanglement for adjacent intervals of size $\ell=50$
in the infinite geometry.
Left: Equal temperatures $\beta_l=\beta_r=\beta$. Horizontal lines indicate the
steady-state entanglement for $\beta=5,10$.
Right: Entanglement evolution for unequal temperatures (symbols)
compared to averages of equal-temperature curves (lines).}
\label{fig:lnft}
\end{figure}

We finally study the case where the initial states on left and right-hand side
are prepared at finite temperatures. We consider again the infinite geometry, where
the initial covariance matrices are given by Eq.~(\ref{c0}) with the sum replaced by an integral.
First, we consider the unbiased case $\beta_l=\beta_r=\beta$, with results on the time
evolution of $\mathcal{E}$ shown on the left of Fig.~\ref{fig:lnft}.
When compared to the local quench results in Fig.~\ref{fig:lnlqi}, one sees that
the curves become flatter and eventually saturate in time for increasing temperatures.
Nevertheless, one observes
the same light-cone effect at $t \approx \ell$ and after a sudden decrease $\mathcal{E}$
relaxes slowly towards its thermal-state value (\ref{lnbeta}).

The case of unequal temperatures is shown on the right of 
Fig.~\ref{fig:lnft}. The result
(\ref{lnbetalr}) for the steady state suggests, that the same relation might be true
for the time evolution as well. Indeed, the average of the time evolved entanglement
with equal temperature initial conditions, $\mathcal{E}(\beta_l)$ and $\mathcal{E}(\beta_r)$,
is indistinguishable from the data $\mathcal{E}(\beta_l,\beta_r)$ for unequal temperatures.
There are, however, again some small deviations.

%%%%%%%%%%%%%%%%%%%%%%%%%%%%%%%%%%%%%%%%%%%%%%%%%%%%%%%%%%%
%
%
%%%%%%%%%%%%%%%%%%%%%%%%%%%%%%%%%%%%%%%%%%%%%%%%%%%%%%%%%%%

\section{Discussion}

In conclusion, we have studied the entanglement, measured by the logarithmic negativity,
in a simple steady state of the harmonic oscillator chain driven by thermal reservoirs at
different temperatures. The steady-state density matrix factorises into two
Gibbs-like states, with Hamiltonians given by only the left- or right-moving particles, and the
entanglement is found to be the sum of the chiral contributions. These are 
simply equal to half of the thermal-state entanglement corresponding to each reservoir,
which can be found by CFT calculations.

The above additivity property depends crucially on the assumption that the effect
of the zero-mode, which couples the chiral branches, can be neglected.
This seems not to be valid for the steady state of a free-fermion chain,
prepared with the same initial condition as considered here. Indeed, in the latter case the 
mode-occupation $n_q$,
given by the  Fermi-statistics analogue of Eq.~(\ref{nq}), develops a jump singularity
around $q=0$. This leads to a contribution in the mutual
information of two adjacent segments, which scales logarithmically in the segment size \cite{EZ14, AEFS14},
and thus the additivity of the chiral contributions does not hold for this measure.
Note, however, that in the CFT limit $\beta_l,\beta_r \gg 1$, the prefactor of the logarithm
is exponentially suppressed as a function of the inverse temperatures,
and numerical
results indicate that the additivity is practically recovered even for very large 
subsystem sizes.

In the opposite limit of high temperatures, the additivity property for the logarithmic
negativity is weakly violated for the harmonic chain, however, the area law is still 
strictly obeyed.
The question thus remains open, whether the logarithmic violation of the area law 
found for the mutual information in the fermionic NESS could persist for the entanglement negativity.
As a further extension, one could test the additivity property of the entanglement for interacting
models, such as the NESS of the XXZ chain \cite{KIM13} or of special integrable quantum field
theories \cite{CCDH14}.

\ack{
The work of V.E. was realized in the framework of T\'AMOP 4.2.4.A/1-11-1-2012-0001
``National Excellence Program''. The project was supported by the European Union
and co-financed by the European Social Fund. Z.Z. acknowledges funding by 
the British Engineering and Physical Sciences Research Council (EPSRC), the Basque
Government (Project No. IT4720-10), and by the European Union through the
ERC Starting Grant GEDENTQOPT and the CHIST-ERA QUASAR project.
}

%\section{Appendix}
%\subsection{Covariance matrices with Dirichlet boundary conditions}

\appendix

\section{Steady-state covariance matrix
\label{app:sscov}}

In this Appendix we derive the NESS covariance matrix in Eq. (\ref{cinf}), following 
 the steps of Ref.~\cite{BPT83}
and correcting a factor of 2 error there. The main idea of the calculation is that the only relevant information
about the initial state, stored in $C_{m,n}(0)$, which survives in the limit $t \to \infty$ of the time evolution is
located asymptotically far away from the initial cut. There the covariances are translationally invariant and given by
\eq{
\lim_{m,n \to \pm \infty} C_{m, n}(0) = \sigma^{\alpha}_{m,n} = 
\int_{-\pi}^{\pi} \frac{\dd q}{2\pi} \ee^{iq(m-n)}
\twomat{\omega_{q}^{-1}}{0}{0}{\omega_{q}}
\coth \frac{\beta_{\alpha} \omega_{q} }{2},
\label{limpm}}
where the $+$ ($-$) sign in the limit corresponds to the right (left) hand side with $\alpha = r$ ($\alpha = l$).
The Fourier transform of $\sigma^{\alpha}_{m,n}$ is denoted by $\sigma^{\alpha}_q$ and we introduce the
notation $\sigma^{\pm}_q = (\sigma^r_q \pm \sigma^l_q)/2$.

Now we split up the initial covariance matrix in three terms as
$\Cov_{m,n} (0)= \Cov_{m,n}^{1}(0)+ \Cov_{m,n}^{2}(0) + \Cov_{m,n}^{3}(0)$ where
\eq{
\Cov^{1}_{m,n}(0)=\frac{\sigma^{r}_{m,n} + \sigma^{l}_{m,n}}{2},\quad
\Cov^{2}_{m,n}(0)=\mathrm{sgn}(m)\frac{\sigma^{r}_{m,n} - \sigma^{l}_{m,n}}{2},
}
and $\Cov_{m,n}^{3}(0)$ describes the remaining terms. Note, that the sum of the first
two terms gives the correct asymptotic behaviour in (\ref{limpm}), however, they generate
nonzero matrix elements in the offdiagonals of $C(0)$ in Eq.~(\ref{ct}), which have to be
compensated by $\Cov_{m,n}^{3}(0)$. The time evolved matrices are given by
$\Cov^{i}(t) = S(t) \Cov^{i}(0) S(t)^T$ such that $\Cov(t)=\Cov^{1}(t) + \Cov^{2}(t) + \Cov^{3}(t)$.

First, we consider $C^1(t)$ which is a product of three Toeplitz matrices and thus
the multiplication can be performed in Fourier space $C^1_q(t) = S_q(t) \sigma^+_q S^{T}_q(t)$.
The symbol of the matrix in Eq.~(\ref{sti}) can be rewritten as a sum of diagonal 
and offdiagonal contributions
\eq{
S_q(t) = \cos (\omega_q t) \, \unity + \sin (\omega_q t) \, \Gamma_q, \qquad
\Gamma_q = \twomat{0}{\omega_q^{-1}}{-\omega_q}{0}.
}
Multiplying out $C^1_q(t)$ and taking $t \to \infty$, the rapidly oscillating terms can be
substituted by their average and we arrive at
\eq{
\lim_{t \to \infty} C^{1}_{m,n}(t)= \int_{-\pi}^{\pi}\frac{\dd q}{2\pi} \ee^{iq(m-n)}
\frac{1}{2} \left(\sigma^+_q + \Gamma_q \sigma^+_q \Gamma^T_q \right) ,
}
which gives the first term in the integral of Eq.~(\ref{cinf}).

The next step is to obtain the asymptotics of $C^2(t)$. Working again in Fourier space, one
has now to include the Fourier transform of the sign function in $C^2_{m,n}(0)$ which leads to
\eq{
C^{2}_{m,n}(t) = \int_{-\pi}^{\pi}  \frac{\dd q}{2\pi} \ee^{iq(m-n)}
P \int_{-\pi}^{\pi} \frac{\dd p}{2\pi} \ee^{-i(p-q)n}
\frac{2i}{p-q} S_q(t) \sigma^-_p S^T_p(t) ,
}
where the $P$ sign indicates that the Cauchy principal value has to be taken. It can be
evaluated using the identity
\eq{
\lim_{t\to\infty} P \int_{-\pi}^{\pi} \frac{\dd p}{p-q} g(p) \ee^{if(p)t} =
i \pi g(q) \ee^{if(q)t} \, \mathrm{sgn} \left[ f'(q) \right],
\label{pint}}
where $f$ and $g$ denote some smooth functions. In turn, the principal value integral has the
effect of interchanging the oscillatory terms $\cos(\omega_p t) \to -\sin(\omega_q t)$
and $\sin(\omega_p t) \to \cos(\omega_q t)$ in $S^T_p(t)$. In the remaining integral
over $q$, one can again take the averages of the time dependent terms which yields
\eq{
\lim_{t\to\infty}  C^{2}_{m,n}(t)= \int_{-\pi}^{\pi}\frac{\dd q}{2\pi}
\ee^{iq(m-n)} i \, \mathrm{sgn}(\omega'_q)
\frac{1}{2} \left( \sigma^-_q \Gamma^T_q - \Gamma_q \sigma^-_q \right) .
}
Noting that in our case $\mathrm{sgn}(\omega'_q) = \mathrm{sgn}(q)$ 
 and calculating the matrix
products leads to  the second term in Eq.~(\ref{cinf}).

We finally show that $\lim_{t \to \infty} \Cov^{3}(t)=0$. Using the form of $C(0)$ in
Eqs.~(\ref{ct}) and (\ref{c0}) as well as the expression for the eigenvectors in (\ref{phik}),
one has in the thermodynamic limit ($N\to\infty$) the block-matrix form
\eq{
C^3(0) =
 -\twomat{\tilde \sigma^l}{\sigma^l}{\sigma^r}{\tilde \sigma^r}, \qquad
\tilde \sigma^\alpha_{m,n} = \int_{-\pi}^{\pi}\frac{\dd q}{2\pi} \ee^{iq(m+n)} \sigma_q^\alpha \, .
%-\theta(m)\theta(n) \tilde \sigma^r_{m,n} - \theta(-m)\theta(-n) \tilde \sigma^l_{m,n}
%-\theta(m)\theta(-n) \sigma^r_{m,n} - \theta(-m)\theta(n) \sigma^l_{m,n}
% - \theta^{++}_{m,n} \tilde \sigma^r_{m,n} - \theta^{--}_{m,n} \tilde \sigma^l_{m,n}
% - \theta^{+-}_{m,n} \sigma^r_{m,n} - \theta^{-+}_{m,n} \sigma^l_{m,n}
%
}
%
%where $\theta(n)=\left[1+\mathrm{sgn}(n)\right]/2$
%
The $\tilde \sigma^\alpha$ in the diagonal are Hankel matrices (with symbol $\sigma^\alpha_q$)
that arise due to the Dirichlet boundary condition in the center, while the offdiagonal Toeplitz matrices
$\sigma^{\alpha}$ compensate the extra contributions from $C^1(0)+C^2(0)$, as mentioned before.
The time-evolved matrix $C^3(t)$ has thus four different contributions from the various Hankel/Toeplitz
blocks which can be written as
\eq{
\fl
C^3_{m,n}(t)= \sum_{\alpha=l,r} \sum_{s=\pm}
\int_{-\pi}^{\pi}\frac{\dd q}{2\pi} P \int_{-\pi}^{\pi}\frac{\dd p}{2\pi} \int_{-\pi}^{\pi}\frac{\dd p'}{2\pi}
\ee^{i(pm-p'n)} \mathcal{F}^\alpha_s(q,p,p') S_p(t) \sigma^{\alpha}_q S^T_{p'}(t),
\label{c3t}}
with the definition
\eq{
\mathcal{F}^r_{\pm}(q,p,p') = 
\frac{1}{2}\left[2\pi\delta(p-q) - \frac{2i}{p-q}\right]
\frac{1}{2}\left[2\pi\delta(p' \pm q) \pm \frac{2i}{p' \pm q}\right],
}
and $\mathcal{F}^l_{\pm}(q,p,p') = \left[\mathcal{F}^r_{\pm}(q,p',p)\right]^*$. Note, that $\mathcal{F}^r_\pm$
is just the Fourier transform of the product of step functions
$\left[1+\mathrm{sgn}(m)\right]/2 \times \left[1\pm\mathrm{sgn}(n)\right]/2$ which singles out
the lower right/left block. Carrying out the integrals over $p$ and $p'$ in Eq.~(\ref{c3t}) 
using formula (\ref{pint}), one arrives at
\begin{eqnarray}
\lim_{t\to\infty} P \int_{-\pi}^{\pi}\frac{\dd p}{2\pi} \int_{-\pi}^{\pi}\frac{\dd p'}{2\pi}
\ee^{i(pm-p'n)} \mathcal{F}^r_\pm(q,p,p') S_p(t) \sigma^r_q S^T_{p'}(t) = 
\nonumber \\
\fl
\frac{\ee^{iq(m \pm n)}}{4}
\left[ S_q(t) \sigma^r_q S^T_q(t) - \hat S_q(t) \sigma^r_q \hat S^T_q(t)
% \pm \mathrm{sgn}(\omega'_q) \mathrm{sgn}(\omega'_{\mp q})
- i \, \mathrm{sgn}(\omega'_ q)
(\hat S_q(t) \sigma^r_q S^T_q(t) + S_q(t) \sigma^r_q \hat S^T_q(t))
\right]
\end{eqnarray}
where we defined $\hat S_q(t) = -\sin(\omega_q t) \unity + \cos(\omega_q t)\Gamma_q$
and used the symmetry properties $S_{-q}(t) = S_q(t)$, $\hat S_{-q}(t) = \hat S_q(t)$ and
$\mathrm{sgn}(\omega'_{-q}) = -\mathrm{sgn}(\omega'_{q})$. Calculating the matrix products,
one finds that the expression in the brackets vanishes identically. The same holds for
the integrals with $\mathcal{F}^l_{\pm}$ which concludes our proof.

\section{CFT results for local quench
\label{app:cft}}

Here we briefly summarize the method used to obtain $\mathcal{E}$ for a local quench.
For a detailed discussion of ground-state entanglement negativity within CFT,
we refer to Ref.~\cite{CCT13}. 

The essential step is to rewrite Eq.~(\ref{lnrho}) as
\eq{
\mathcal{E} = \lim_{n_e \to 1} \ln \Tr (\rho_A^{T_2}(t))^{n_e}
\label{lnrepl}}
where $n_e$ is an even integer which, at the end of the calculation, has to be analytically
continued to one. Thus, one
first needs to express the trace of an even power of the partial-transposed reduced density matrix
$\rho_A(t) = \Tr_B \rho(t)$ in terms of a path integral. This is done by considering
an $n_e$-sheeted Riemann surface and sewing together the replicas of $\rho_A^{T_2}(t)$.
Each copy can be represented by a 2D path integral with slits along the intervals
$A_1$ and $A_2$ on the real axis, and partial transposition $T_2$ corresponds to interchanging
the edges of the corresponding slit $A_2$.

Instead of working on a replicated world-sheet, one can introduce replicated fields and
the so-called \emph{twist fields}, $\mathcal{T}_{n_e}$ and $\overline{\mathcal{T}}_{n_e}$, that
cyclically permute replicas in one of two directions. Each time when replicas are sewn together,
one has to calculate expectation values of products of the two twist fields, inserted at the
endpoints of the slits. Whenever the slit corresponds to the partial transposed subsystem,
the twist operators have to be interchanged.

%%%%%%%%%%%%%%%%%%%%%%%%%%%%%%%%%%%%%%%%%%%%%%%%%%%%%%%%%%%
%
\begin{figure}[htb]
\center
\includegraphics[width=0.5\columnwidth]{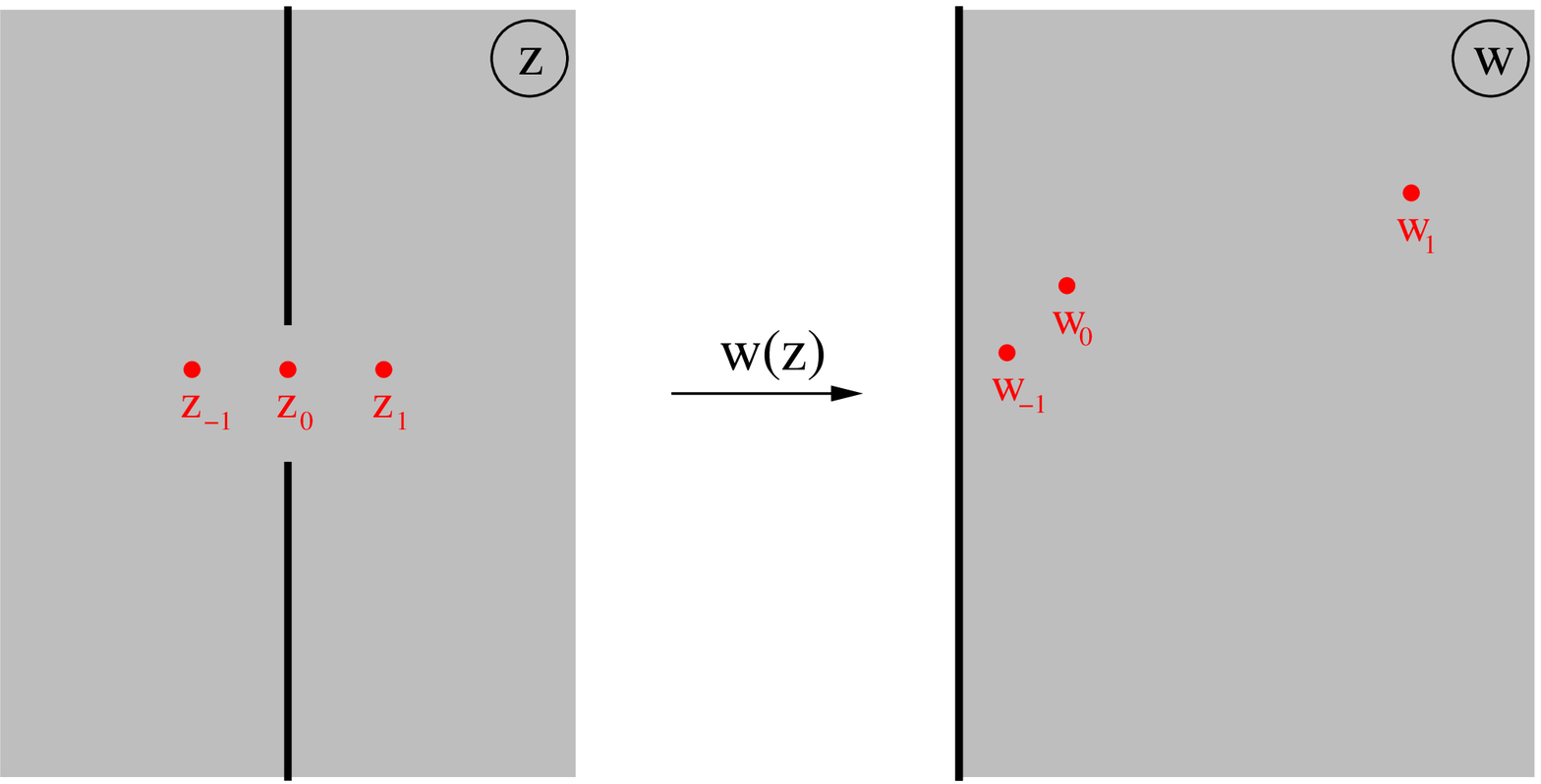}
\caption{Conformal map from the quench geometry to the half-plane. Red dots indicate
the locations of twist-field insertions and their images under $w(z)$.}
\label{fig:confmap}
\end{figure}
%
%%%%%%%%%%%%%%%%%%%%%%%%%%%%%%%%%%%%%%%%%%%%%%%%%%%%%%%%%%%

The simplest choice for the subsystems is a bipartition ($B=\emptyset$) into two semi-infinite lines
$A_1$ and $A_2$. One has then a single contact point and thus
\eq{
\Tr (\rho_A^{T_2}(t))^{n_e} = \langle \mathcal{T}^2_{n_e}(z_0) \rangle
\label{1p}}
where the expectation value of the twist fields has to be calculated on a world-sheet representing
the time-evolved density operator $\rho(t)$. This is depicted on the left of 
Fig.~\ref{fig:confmap}, where
the cut in the middle corresponds to the imaginary time evolution of two decoupled CFTs with fixed
boundary conditions, yielding the initial state $\rho(0)$. The real time evolution takes place between the
endpoints $z=\pm i\epsilon$ of the cut, and the twist field has to be inserted at $z_0=i \tau$.
Note, that the parameter $\epsilon$ is needed to regularize the path-integral and the analytical
continuation to real times $\tau = it$ must be carried out at the end of the calculation.

Although the original world-sheet has a complicated geometry, one can apply a conformal
mapping \cite{CC07}
\eq{
w = \frac{z}{\epsilon} + \sqrt{\left(\frac{z}{\epsilon}\right)^2 + 1}
\label{confmap}}
which transforms it to the half-plane, as shown on the right of 
Fig.~\ref{fig:confmap}. On the
half-plane the expectation value of a one-point function is known
and one can then use the conformal transformation formula to obtain
\eq{
\langle \mathcal{T}^2_{n_e}(z_0) \rangle \propto \left(
\left|\frac{\dd w}{\dd z}\right|_{z_0} \frac{1}{\mathrm{Re}(w_0)} \right)^{\Delta_{n_e}}
\label{1pct}}
where $w_0=w(z_0)$ and the scaling dimension of the operator $\mathcal{T}^2_{n_e}$
is given by \cite{CCT13}
\eq{
\Delta_{n_e} = \frac{c}{6}\left( \frac{n_e}{2} - \frac{2}{n_e}\right)
\label{sd}}
with the central charge $c$. Carrying out the calculations, one obtains
\eq{
%\langle \mathcal{T}^2_{n_e}(z_0) \rangle \propto
\Tr (\rho_A^{T_2}(\tau))^{n_e} \propto
\left( \frac{\epsilon}{\epsilon^2 - \tau^2} \right)^{\Delta_{n_e}}
\label{1pres}}
Continuing the result to real time $\tau \to it$, taking the limit $t \gg \epsilon$ and finally
using Eq.~(\ref{lnrepl}), one arrives to the result in Eq.~(\ref{lnlqh}).

The local quench for the finite system can be treated in a similar way. There the world-sheet
has a double-pants geometry, with fixed boundary conditions along $\mathrm{Re}(z)=\pm \ell$,
and the proper conformal mapping to the half-plane was given in Ref.~\cite{SD11}.
Carrying out the analogous steps, one arrives to the formula in Eq.~(\ref{lnlqf}).

On the other hand, the tripartite case in the infinite system with line segments $A_1=\left[-\ell,0\right]$
and $A_2=\left[0,\ell \right]$ is more involved. There, one has to consider the three-point function
$\langle \mathcal{T}(z_{-1}) \overline{\mathcal{T}}^2_{n_e}(z_0)  \mathcal{T}(z_{1})\rangle$ of the twist fields with
$z_0=i\tau$ and $z_{\pm 1}=\pm \ell+i\tau$. This is mapped under (\ref{confmap}) into a three-point function on
the half-plane which, however, has the complexity of a six-point function on the full plane.
Although some recent progress has been made in the derivation of higher order twist-field correlators in Ref.~\cite{CTT14}, 
their structure is rather involved and we have not been able to tackle this case analytically.

\section*{References}

\bibliographystyle{iopart-num}

\bibliography{oscneg_refs}

\end{document}